\documentclass[english,twocolumn,prl,amssymb,amsmath,superscriptaddress]{revtex4}
\usepackage[latin9]{inputenc}
\usepackage{graphicx,color}
\usepackage{hyperref}
\usepackage{babel}
\usepackage[normalem]{ulem}

\newcommand{\be}{\begin{equation}}
\newcommand{\ee}{\end{equation}}

\definecolor{mygreen}{rgb}{0,0.5,0}
\definecolor{myblue}{rgb}{0,0,0.75}
\definecolor{mymagenta}{cmyk}{0,1,0,0.12}

\begin{document}

\title{Dynamical Topological Quantum Phase Transitions for Mixed States}

\author{M. Heyl}
\affiliation{Max-Planck-Institut f\"ur Physik komplexer Systeme, N\"othnitzer Stra\ss e 38, 01187-Dresden, Germany}

\author{J. C. Budich}
\affiliation{Department of Physics, University of Gothenburg, SE 412 96 Gothenburg, Sweden}
	
\date{\today}

\begin{abstract}

We introduce and study dynamical probes of band structure topology in the post-quench time-evolution from mixed initial states of quantum many-body systems. Our construction generalizes the notion of dynamical quantum phase transitions (DQPTs), a real-time counterpart of conventional equilibrium phase transitions in quantum dynamics, to finite temperatures and generalized Gibbs ensembles. The non-analytical signatures hallmarking these mixed state DQPTs are found to be characterized by observable phase singularities manifesting in the dynamical formation of vortex-antivortex pairs in the interferometric phase of the density matrix. Studying quenches in Chern insulators, we find that changes in the topological properties of the Hamiltonian can be identified in this scenario, without ever preparing a topologically non-trivial or low-temperature initial state. Our observations are of immediate relevance for current experiments aimed at realizing topological phases in ultracold atomic gases. 

\end{abstract}

\date{\today}

\maketitle

\emph{Introduction. }
The study of topological states of matter has been a major focus of research in physics for many years.
Theoretical \cite{Jaksch2003,Duan2006,GerbierDalibard,DalibardReview,Cooper2011,Cooper2013}
and experimental \cite{Aidelsburger2011,Sengstock,Ketterle2013,Aidelsburger2013,jotzu2014,Atala2014,Aidelsburger2015} progress on realizing such exotic phases in ultracold atomic gases (see Ref. \cite{NathanReview} for a review) now provides a platform to study their nonequilibrium properties.
In particular, the latter systems naturally offer powerful probes beyond what is achievable in solid state materials, including full state tomography \cite{HamburgStateReconstruction} and single-site resolved quantum gas microscopy \cite{GreinerMicroscope,Weitenberg:2011,Endres,GreinerEntanglement}.
A generic protocol in state of the art cold atom experiments is to prepare a topologically trivial (thermal) initial state and then observe the non-equilbrium dynamics after the Hamiltonian has been quenched into a topological phase.
In this scenario, a natural challenge is to reveal dynamical signatures of topology, without ever preparing a low-temperature state of the final target Hamiltonian \cite{CooperQuenchCI, YingDynamicalHall, RefaelDynamicalHall,WangNoneqChern}.

\begin{figure}[htp]
\centering
\includegraphics[width=\columnwidth]{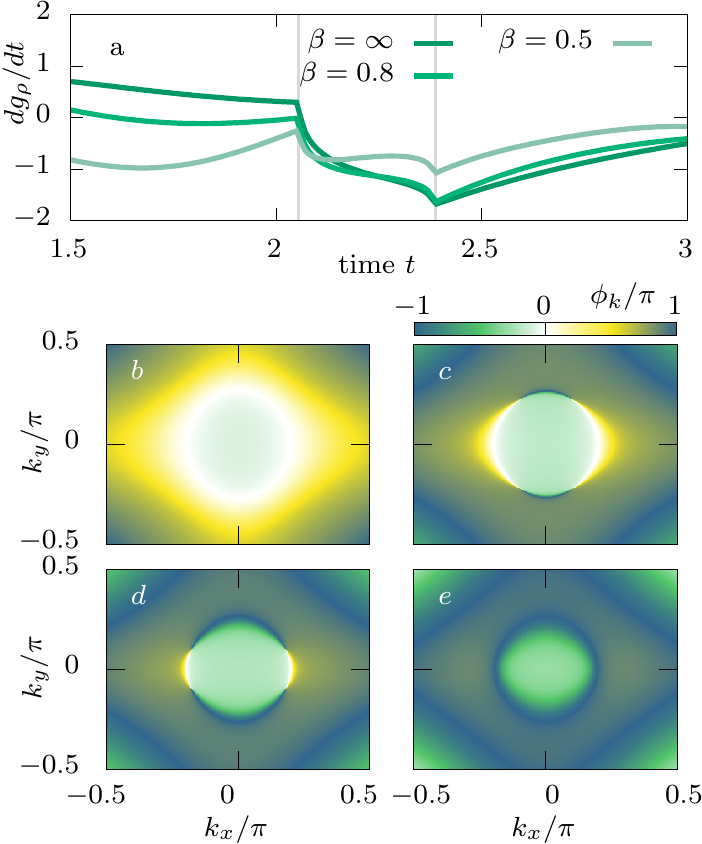}
\caption{Panel (a): Dynamical quantum phase transitions (DQPTs) in the Chern insulator Dirac model (see Eq. (\ref{eqn:DiracCI})) for initial states at different temperatures. The DQPTs are visible as cusps at the temperature-independent critical times $t^\ast_c$ and $t^\ast_a$ (marked with grey lines) in the time-derivative of the Loschmidt rate function $dg_\rho/dt$. We show data for a quenches from parameters $m_i=5.0, \lambda_i=3.0$ to a final $m_f=1.5,\lambda_f=1.0$ at several initial inverse temperatures $\beta$. Panels (b-e): Momentum-dependent phase of $\mathcal G_\rho^k(t)$ at $t=1.5<t^\ast_c$ (b), $t^\ast_c<t=2.1<t^\ast_a$ (c), $t^\ast_c<t=2.3<t^\ast_a$ (d), and $t=2.6>t^\ast_a$ (e), exhibiting phase vortices created at $t^\ast_c$ and annihilated at $t^\ast_a$.}
\label{fig:dqptMDM}
\end{figure}

The purpose of this work is to propose and analyze dynamically defined probes of such topologically non-trivial quenches for realistic systems described by a mixed state.
To this end, we generalize from pure states to density matrices the concept of dynamical quantum phase transitions (DQPTs) \cite{Heyl2013a} which defines a nonequilibrium counterpart to conventional transitions in quantum real-time evolution.
While at an equilibrium transition the thermodynamic potentials become nonanalytic as a function of a control parameter such as temperature, DQPTs lead to non-analytic behavior as a function of time in Loschmidt amplitudes $\mathcal G(t)=\langle \psi(0)\vert \psi(t)\rangle$, quantifying the time-dependent deviation from the initial state $|\psi(0)\rangle$.
Recently, DQPTs have been observed experimentally both with ultra-cold atoms~\cite{HamburgExperiment} and trapped ions~\cite{IonExperiment}.

\emph{Generalized Loschmidt echo and mixed state DQPTs. } Here, we study a generalization of the Loschmidt-Echo to general density matrices defined as
\begin{align}
\mathcal G_\rho(t) = \text{Tr}[\rho_0U(t,0)],
\label{eqn:Loschmidtdens}
\end{align} 
where $\rho_0$ is the density matrix describing the initial state and $U(t,0)$ is the time evolution operator. The relation between $\mathcal G_\rho$ and the standard Loschmidt echo becomes manifest when purifying $\rho_0=\sum_j w_j \lvert j\rangle\langle j\rvert$ as $\lvert\Psi^\rho_0\rangle=\sum_j \sqrt{w_j}\lvert j\rangle\otimes\lvert j\rangle_a$, where $\lvert j\rangle_a$ are orthonomal states in an ancilla Hilbert space $\mathcal H_a$ that is chosen isomorphic to the physical Hilbert space $\mathcal H$. With the time evolved purified state $\lvert \Psi^\rho(t)\rangle = U(t,0)\otimes \mathbf 1_a\lvert \Psi_0^\rho\rangle$, the relation 
\begin{align}
\mathcal G_\rho(t)=\langle \Psi_0^\rho\vert \Psi^\rho(t)\rangle
\label{eqn:PurifiedLoschmidt}
\end{align}
establishes the formal equivalence to the standard Loschmidt echo $\mathcal G(t)$ for pure states. In complete analogy to the pure state case, the Fourier transform of $\mathcal G_\rho$ is the power distribution $P(E)$ of the system's energy $E$ after the quench.

We note that the extension of the Loschmidt amplitude to mixed states is not unique, as several generalizations that are consistent with the pure state case are conceivable (see e.g. Ref. \cite{AbelingKehrein} for an alternative definition). Our present definition (\ref{eqn:Loschmidtdens}) is motivated by the recent insight \cite{BudichHeyl2016}  that DQPTs are closely connected to discontinuities in geometric phases \cite{Pphase, PphasePRL}. Replacing the role of the pure state geometric phase by the interferometric phase for mixed states \cite{InterferometricPhase} naturally leads us to Eq.~(\ref{eqn:Loschmidtdens}) which, at a formal level, is further corroborated by Eq. (\ref{eqn:PurifiedLoschmidt}).

While thermal phase transitions are accompanied by non-analyticities in the free energy density, DQPTs are hallmarked by non-analytic behavior of the rate function $g_\rho(t)$ of the Loschmidt amplitude (see e.g. Fig.~\ref{fig:dqptMDM}) which in our mixed-state context is defined as
\begin{align}
g_\rho(t)=-\frac{1}{N}\log\lvert\mathcal G_\rho(t)\rvert^2,
\label{eqn:raterho}
\end{align}
where $N$ is the number of degrees of freedom of the system. In the pure state limit, it has been shown that DQPTs inherit many essential properties of conventional equilibrium phase transitions including robustness against weak perturbations~\cite{Karrasch,Kriel,Canovi,Sharma} or scaling and universality~\cite{Scaling}.

Analogous to the pure state case, non-analyticities in $g_\rho(t)$ are closely connected to Fisher-zeros in $\mathcal G_\rho(t)$, where the time evolved purified state $\lvert\Psi^\rho(t)\rangle$ becomes orthogonal to the initial state $\lvert \Psi_0^\rho\rangle =\lvert\Psi^\rho(t=0)\rangle$ (see Eq. (\ref{eqn:PurifiedLoschmidt})). For the standard Loschmidt amplitude $\mathcal G(t)$, it has been shown \cite{BudichHeyl2016} that Fisher zeros are always accompanied with $\pi$-phase slips of the Pancharatnam geometric phase \cite{Pphase,PphasePRL}. Here, we find that Fisher-zeros in $\mathcal G_\rho(t)$ are directly related to such phase jumps in the interferometric phase introduced in Ref. \cite{InterferometricPhase} to describe geometric effects in interference experiments on mixed states.

\emph{Signatures of topology in quench dynamics. }
To reveal the role of topology in mixed state DQPTs and their potential as dynamical probes of topological properties, we study sudden quantum quenches from an initial Hamiltonian $H_i$ to a final Hamiltonian $H_f$ at time $t=0$. In contrast to earlier work assuming the initial state to be the ground state of the initial Hamiltonian, here we only require the initial density matrix $\rho_0$ to be in a generalized Gibbs ensemble with respect to $H_i$, i.e. to be diagonal in the basis of $H_i$.

The class of Hamiltonians we consider are gapped band structures encompassing insulating fermionic systems within the independent particle approximation and superconducting systems at the mean-field level, respectively. For concreteness, we illustrate our construction with two-band models following recent experiments in ultracold atoms~\cite{Aidelsburger2011,Sengstock,Ketterle2013,Aidelsburger2013,jotzu2014,Atala2014,Aidelsburger2015}. We note, however, that the extension of DQPTs to multi-band systems has been generally analyzed in Ref. \cite{BalatskyMultiBand}. Turning to momentum space, the Hamiltonians of interest can be cast in the form
\begin{align}
h_k^\mu=\vec d_k^\mu \cdot \vec \sigma, \quad \mu=i,f,
\label{eqn:hamiltonian}
\end{align}
where $\vec \sigma=(\sigma^x,\sigma^y,\sigma^z)$ denotes the vector of standard Pauli matrices and $\vec d_k^i$ ($\vec d_k^f$) is a real three-component vector characterizing the initial (final) Bloch Hamiltonian at lattice momentum $k$. This choice of representation of the Hamiltonian automatically fixes the zero of energy. The Bloch eigenstates of the upper ($+$) and lower ($-$) bands before ($\mu=i$) and after ($\mu=f$) the quench are denoted by $\lvert \mu_k^\pm\rangle$ and the corresponding energy eigenvalues are $\epsilon_{k,\pm}^\mu=\pm\epsilon_k^\mu$. In this notation, the initial density matrix takes the form 
\begin{align}
\rho_k(0)=\frac{1}{2}\left(1-\vec n_k(0) \cdot \vec \sigma \right)=p_k \lvert i_k^-\rangle\langle i_k^-\rvert + (1-p_k)\lvert i_k^+\rangle\langle i_k^+\rvert,
\label{eqn:rhoinmomentum}
\end{align}
where $\vec n_k(0)$ denotes the initial Bloch vector and the probability $0<p_k\le 1$ of being in the lower band of the initial Hamiltonian parameterizes the translation-invariant generalized Gibbs state.

To explicitly compute the density matrix Loschmidt amplitude $\mathcal G_\rho(t)=\prod_k\mathcal G_\rho^k(t)$, we expand the initial eigenstates in the basis of the final Hamiltonian as $\lvert i_k^-\rangle = g_k \lvert f_k^-\rangle + e_k \lvert f_k^+\rangle$ and $\lvert i_k^+\rangle= e_k^*\lvert f_k^-\rangle-g_k^*\lvert f_k^+\rangle$, respectively. In this notation, the central quantity $\mathcal G_\rho^k(t)$ is readily expressed explicitly as
\begin{align}
\mathcal G_\rho^k(t)= & p_k(\lvert g_k\rvert^2\text{e}^{i\epsilon_k^f t}+\lvert e_k\rvert^2\text{e}^{-i\epsilon_k^f t})+\nonumber\\ 
&(1-p_k)(\lvert e_k\rvert^2\text{e}^{i\epsilon_k^f t}+\lvert g_k\rvert^2\text{e}^{-i\epsilon_k^f t}).
\label{eqn:LoMixedMomentum}
\end{align}
For pure initial states, i.e. for $p_k\equiv 1$, it is well known \cite{VajnaDora,BudichHeyl2016} that Fisher zeros only occur at so called critical momenta $k_c$ which are characterized by $\lvert g_{k_c}\rvert^2=\lvert e_{k_c}\rvert^2=1/2$, or equivalently by $\vec d_{k_c}^i\cdot \vec d_{k_c}^f=0$. Remarkably, even for general mixed initial states (see Eq. (\ref{eqn:rhoinmomentum})), we find at critical momenta independently of $p_k$, i.e. in particular independently of temperature for thermal states that
\begin{align}
\mathcal G_\rho^{k_c}(t)=\cos(\epsilon_{k_c}^ft).
\label{eqn:criticalLoschmidt}
\end{align}
According to Eq.~(\ref{eqn:criticalLoschmidt}), the amplitude $\mathcal{G}_\rho^{k_c}(t)$ at a critical momentum $k_c$ changes sign at times $t_c^n=\pi/(2 \epsilon_{k_c}^f)(2n +1),~n\in \mathbb N_0$, leading to a $\pi$-jump of the associated phase $\phi_k(t)$ defined via:
\begin{align}
\mathcal{G}_\rho^k(t) = r_k(t) e^{i\phi_k(t)},
\label{eqn:defPhi}
\end{align}
thus imprinting a unique signature of DQPTs in the phase profile $\phi_k(t)$. Importantly,  $\phi_k(t)$ is accessible experimentally using interferometry for mixed states~\cite{InterferometricPhase}. Furthermore, as our subsequent analysis shows, mapping out the time-dependent Bloch-vector $\vec n_k(t)$ (see Eq. (\ref{eqn:rhoinmomentum})) along the lines of Refs. \cite{HamburgStateReconstruction,HamburgExperiment} also allows to detect clear signatures of the DQPTs discussed here.

From Eq.~(\ref{eqn:criticalLoschmidt})  we conclude that Fisher-zeros and, as a consequence DQPTs, occur in all situations where they would have occurred for pure initial states at the same critical times. Furthermore, it is easy to see from Eq. (\ref{eqn:LoMixedMomentum}) that as long as $p_k>0.5$ for all $k$, including in particular all finite temperature states, no additional Fisher-zeros can arise compared to the pure state case. This provides a strong generalization of the notion of DQPTs to mixed states \cite{foot1}. In contrast, if $p_k=0.5$, we always find $\mathcal G_\rho^{k}(t)=\cos(\epsilon_{k}^ft)$ irrespective of the values of $g_k$ and $e_k$ which renders infinite temperature a singular point with Fisher-zeros at all momenta.

\begin{figure}
\centering
\includegraphics[width=\columnwidth]{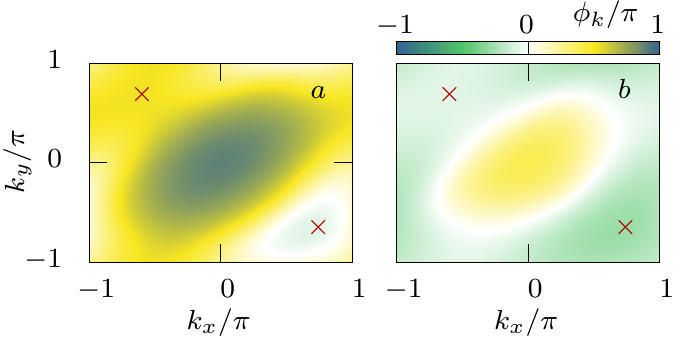}
\caption{Critical contours for the Haldane model with $\phi=\pi/2$, $V=1$, and $V'=1/4$ at nonzero initial temperature $\beta = 1$. The red crosses mark the locations of the Dirac points of the two topological transitions at $M_c = \pm 3\sqrt{3}V'$. Panel (a): Phase profile at time $t=1/2$ for a topologically-protected DQPT with a quench from $m_i=2$ to $m_f=1/2$ where the critical contour, along the white line here, encloses the Dirac point of the underlying crossed topological transition. Panel (b): Phase profile at time $t=0.4$ for a quench from $m_i=2$ to $m_f=-2$, starting and ending in the topologically trivial phase. Here, the critical contour separates two regions in the Brioullin zone either incorporating both Dirac points or none.}
\label{fig:phaseHaldane}
\end{figure}

For pure initial states, a deep relation between band structure topology and the occurrence of DQPTs has been revealed \cite{VajnaDora,BudichHeyl2016,BalatskyMultiBand,HamburgExperiment}: Changes in the topological invariants of the Hamiltonian over the quench imply the occurrence of DQPTs in the post-quench dynamics. Furthermore, DQPTs that are in this sense of topological origin can be phenomenologically distinguished \cite{BudichHeyl2016,HamburgExperiment} from accidental ones that do not involve changes in the band structure topology (see also Fig.~\ref{fig:phaseHaldane} and its discussion below). A central result of our present work is that, quite surprisingly, these sharp non-analytical features in the real time evolution of quantum many-body systems fully survive for a quite general class of mixed states, including states at arbitrary but finite temperature. To illustrate and discuss these insights, we will study two paradigmatic benchmark examples below.

For concreteness, we focus on systems in two spatial dimensions (2D), where the relevant topological invariant characterizing gapped band structures is the integer quantized Chern number $\mathcal C$.  As mentioned before, a change in $\mathcal C$ over the quench implies the presence of critical momenta leading to Fisher zeros and DQPTs. Since $\vec d_k^i$ and $\vec d_k^f$ are smooth functions of $k$, critical momenta hallmarked by $\vec d_{k_c}^i\cdot \vec d_{k_c}^f=0$ occur on closed contours in the first Brillouin zone (BZ) which we call critical contours $\Gamma_c$ in the following. The precise times, for which the $\pi$-phase jumps occur on the contour $\Gamma_c$, depend crucially on the spectrum $\epsilon_c \in [\epsilon_c^\mathrm{min},\epsilon_c^\mathrm{max}]$ corresponding to the critical momenta on $\Gamma_c$, see Eq.~(\ref{eqn:criticalLoschmidt}). The first Fisher zeros occur at the momenta with the highest post-quench energy $\epsilon_{k_c}^f=\epsilon_c^\mathrm{max}$. The associated $\pi$-phase jumps in $\phi_{k_c}(t)$ lead to the formation of vortex anti-vortex pairs in the phase profile $\phi_k(t)$ (see e.g. Fig. \ref{fig:dqptMDM}(c)). Following the energy dispersion on the critical contour, the vortex anti-vortex pairs split and the vortices move along the critical contours until they annihilate with one of their anti-parts at a later time, now determined by the minimum energy $\epsilon_c^\mathrm{min}$ on $\Gamma_c$ (see e.g. Fig.~\ref{fig:dqptMDM}(b-e)).

\emph{Benchmark examples. }As concrete examples we consider two paradigmatic 2D systems, namely the massive Dirac model (MDM) for a Chern insulator on a square lattice \cite{DiracCI}, and the Haldane model \cite{HaldaneModel} on the honeycomb lattice, which has been recently realized in an ultra-cold atom experiment \cite{jotzu2014}. Using the notation of Eq.~(\ref{eqn:hamiltonian}), the MDM is defined as
\begin{align}
\vec d_k = \Big( \lambda \sin(k_x), \sin(k_y), m - \cos(k_x) - \cos(k_y) \Big),
\label{eqn:DiracCI}
\end{align}
with $k=(k_x,k_y)$ denoting the 2D crystal momentum. Here, $m$ represents the Dirac mass and $\lambda>0$ is an anisotropy parameter which allows us to tune the nature of the DQPTs as shown below. Throughout the manuscript we measure length in units of the lattice spacing, and the overall energy scale of the Hamiltonian is set to $1$. The model exhibits three topological phase transitions at $m=\pm 2$ and $m=0$ separating trivial insulating phases for $|m|>2$ from topological phases with $\mathcal{C}=\mathrm{sign}(m)$ for $|m|<2$. We note that varying $\lambda$ only deforms the band structure and affects its symmetry, but leaves the Chern number untouched (given $\lambda>0$).

\begin{figure}
\centering
\includegraphics[width=\columnwidth]{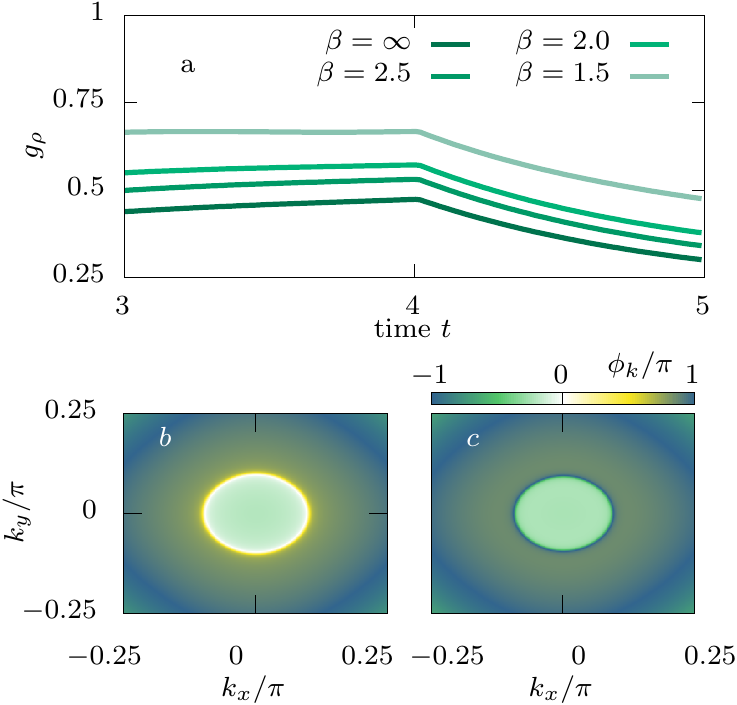}
\caption{DQPTs in the symmetric ($\lambda=1.0$) massive Dirac model (\ref{eqn:DiracCI}). Panel (a): $g_\rho(t)$ for different initial inverse temperatures $\beta$. At the critical time $t^\ast$, $g(t)$ exhibits a kink in the limit $m_i\rightarrow 2+0^+, m_f\rightarrow 2-0^+$, quite visible already for the present quench parameters $m_i=2.3,m_f=1.7$. Panels (b-c): Phase of $\mathcal G_\rho(t)$ for $\beta=1.2, m_i=2.3, m_f=1.7$ at $t=3.9<t^\ast$ (b) and $t=4.1>t^\ast$ (c), respectively. At $t^\ast$ the phase suddenly jumps from $0$ (white) to $\pi$ (blue) on the entire ring-shaped critical contour.}
\label{fig:dqptSymmetric}
\end{figure}

In agreement with the above analysis, topologically protected DQPTs occur at arbitrary temperature $\beta=1/T>0$ initial states, whenever a system is quenched across a topological transition changing the Chern number $\mathcal{C}$ from $0$ to $1$. In Fig.~\ref{fig:dqptMDM} we show numerical data for such a DQPT for a quench from $m_i=5.0,\lambda_i=3.0$ to $m_f=1.5,\lambda_f=1.0$. Panel (a) shows the time derivative of the rate-function $g_\rho(t)$ (see Eq.~(\ref{eqn:raterho}))  at  various inverse temperatures. Remarkably, the temporal non-analyticities at critical times $t^\ast_c$ and $t^\ast_a$ in the form of kinks in $dg_\rho/dt$ hallmarking the DQPTs sharply survive at the same critical times for initial states at arbitrary finite temperature. 

The creation and annihilation of vortex-antivortex pairs in the phase $\phi_k(t)$ is inherently connected with DQPTs in 2D (see Fig. \ref{fig:dqptMDM}), as generally discussed above. Concretely applied to our present example, two vortex-antivortex pairs are created at time $t^\ast_c$ at the top and bottom of the critical contour and annihilated at $t^\ast_a$ after traveling half around the critical contour (Fig. \ref{fig:dqptMDM}(c-e)). The time period between the two DQPTs hence concurs with the lifetime of the vortices.

The length of this temporal interval with nonzero density of Fisher zeros depends crucially on the anisotropy $\lambda$. When $\lambda \to 1.0$, we obtain $t^\ast_a \to t^\ast_c$ in the limit where $m_i$ and $m_f$ approach the critical $2.0$ from opposite sides. In this situation the critical contour lies on the rotationally symmetric Dirac cone that forms around $k=0$ at $m=2.0$. Thus all critical momenta have the same post-quench energy and all corresponding Fisher-zeros collapse to the critical time $t^\ast=t^\ast_a = t^\ast_c$. This leads to a simultaneous singularity in the interferometric phase \cite{InterferometricPhase} on the entire critical contour and thus a kink directly in $g_\rho(t)$ rather than its time derivative. In Fig. \ref{fig:dqptSymmetric}, we illustrate this behavior by plotting $g_\rho(t)$ for various temperatures (panel (a)) as well as the phase of $\mathcal G_\rho^k(t)$ shortly before (panel (b)) and after (panel (c)) $t^\ast$.

As a second benchmark example we consider the Haldane model on a honeycomb lattice~\cite{HaldaneModel}. In the notation of Eq.~(\ref{eqn:hamiltonian}), this model is defined as:
\begin{align}
d_k^x = V\sum_{j=1}^3 \cos(k a_j), \,\, d_k^y = V\sum_{j=1}^3 \sin(k a_j) , \nonumber
\\
d_k^z = m - 2V' \sin(\phi) \sum_{j=1}^3  \cos(k b_j).
\end{align}
For the definition of the vectors $a_j$ and $b_j$ connecting nearest and next-to-nearest neighbor lattice sites on the honeycomb lattice we refer to Ref.~\cite{HaldaneModel}. For $V'/V<1/3$, the system exhibits a trivial insulating phase for $M/V'>3\sqrt{3}|\sin(\phi)|$ and a topological one for $M/V'<3\sqrt{3}|\sin(\phi)|$~\cite{HaldaneModel}.

In Fig.~\ref{fig:phaseHaldane}, we compare topologically protected DQPTs to accidental ones for nonzero initial temperatures. When quenching from the trivial to a non-trivial regime (see Fig.~\ref{fig:phaseHaldane}(a)), we find one critical contour $\Gamma_c$ in $\phi_k(t)$ winding around the Dirac point associated with the topological transition that is crossed by the considered quantum quench. By contrast,  when quenching without changing the Chern number $\mathcal{C}$ (see Fig.~\ref{fig:phaseHaldane}(b)), the model also supports accidental DQPTs if the quench bridges a non-trivial region. However, in this case the critical contour no longer separates the two Dirac points associated with the two crossed topological transitions, but rather separates a region including both Dirac points from a region with none. This provides a clear qualitative distinction between the accidental DQPT and the topologically protected one.

\emph{Concluding discussion. }
The topologically protected mixed state DQPTs introduced in this work provide a powerful tool to dynamically probe changes in the topological properties of the Hamiltonian over a quench in a non-equilibrium fashion, without the necessity of ever preparing a low-temperature state. This is of direct importance for present experiments on ultracold atomic gases, where quenching between topologically distinct Hamiltonians is state of the art, while preparing topologically non-trivial low temperature states remains an open challenge. Full post-quench state reconstruction along the lines of Ref. \cite{HamburgStateReconstruction} provides the necessary tools to observe the implications of our findings in full detail. In particular, by mapping out the time-dependence of the Bloch vector characterizing the density matrix of a two band model, the dynamical vortices hallmarking the mixed state DQPTs discussed in our present work can be fully reconstructed \cite{HamburgExperiment}. 

In our analysis, we have assumed the post-quench dynamics as perfectly coherent. This is a good approximation for the short-time post-quench dynamics considered in the study of DQPTs where the relevant time scales are shorter than coherence times, as confirmed experimentally in optical lattices \cite{HamburgExperiment}. The generalization of our results to open-system dynamics accounting for experimental scenarios with stronger decoherence, as well as taking into account the effect of coupling to higher Bloch bands of the optical lattice are interesting subjects of future work.

Recently, a slightly different approach towards the generalization of DQPTs to nonzero temperatures has been reported \cite{AbelingKehrein}. There, it is found that the non-analyticities characterizing the DQPTs for pure states are smoothed out at nonzero temperature, based on a different extension to mixed states of the Loschmidt amplitude as compared to Eq. (\ref{eqn:Loschmidtdens}). For the considered topological systems, our generalization of DQPTs reproduces the experimental observation of phase vortices in ultracold atoms~\cite{HamburgExperiment}, an experiment unavoidably performed on mixed states.

\emph{Note added. }During the preparation of this manuscript for submission we became aware of a related work on DQPTs for mixed states in 1D systems~\cite{noteadded}.

\emph{Acknowledgements. }We acknowledge discussions with J. Goold, S. Manmana, and L. Pastori, as well as financial support by the Deutsche Forschungsgemeinschaft via the Gottfried Wilhelm Leibniz Prize programme.

\bibliographystyle{apsrev}

\begin{thebibliography}{10}
\bibitem{foot1}{
In the present $2$-band setting, the occurrence of DQPTs is indeed completely independent of temperature. However, at least the independence on the sign of the temperature is due to the equal modulus of the Chern numbers imposed by their zero sum rule, and may thus change in more generic settings with a higher number of bands. 
}
\end{thebibliography}

\end{document}